\documentclass[aps,prd,print,groupedaddress,nofootinbib,eqsecnum,onecolumn]{revtex4}
\usepackage{eurosym}
\usepackage{amsmath}
\usepackage{amsthm}
\usepackage{amssymb}
\usepackage{slashed}
\usepackage{hyperref}
\usepackage{epstopdf}

\usepackage[english]{babel}
\usepackage[latin1]{inputenc}
\usepackage{amsmath, amsthm, amssymb}
\usepackage[pdftex]{graphicx}
\usepackage{amssymb}
\usepackage{latexsym}
\usepackage{amstext}
\usepackage{epstopdf}
\usepackage{floatflt}
\usepackage{hyperref}
\usepackage{enumitem}
\usepackage{multirow}
\usepackage{pdflscape}
\usepackage{makecell}
\usepackage{nopageno}
\newtheorem{theorem}{Theorem}[section]

\newtheorem{definition}[theorem]{Definition}

\newtheorem{remark}[theorem]{Remark}

\newcommand{\aaa}{\mathbf{a}}
\newcommand{\adue}{\aaa_2}

\newcommand{\adc}{\aaa_2^c}

\newcommand{\adf}{\aaa_2^f}

\newcommand{\adj}{\text{ad}}

\newcommand{\as}{\alpha}

\newcommand{\ba}{\begin{array}}

\newcommand{\bb}{\beta}

\newcommand{\be}{\begin{equation}}
\newcommand{\bea}{\begin{equation}\begin{array}}
\newcommand{\beal}{\begin{aligned}}
\newcommand{\beas}{\begin{equation*}\begin{array}}
\newcommand{\bef}{\begin{flalign}}
\newcommand{\befs}{\begin{flalign*}}
\newcommand{\bes}{\begin{equation*}}

\newcommand{\bit}{\begin{itemize}}
\newcommand{\blms}{{\mathfrak B}_{LS}}

\newcommand{\cc}{\mathbf{C}}

\newcommand{\ddn}{\mathbf {d_N}}

\newcommand{\dotto}{\mathbf{d_8}}

\newcommand{\dqn}{\mathbf{d_{4n}}}
\newcommand{\dset}{\mathbf{d_7}}
\newcommand{\ea}{\end{array}}
\newcommand{\eal}{\end{aligned}}

\newcommand{\ee}{\end{equation}}
\newcommand{\eea}{\end{array}\end{equation}}
\newcommand{\eeas}{\end{array}\end{equation*}}
\newcommand{\eef}{\end{flalign}}
\newcommand{\eefs}{\end{flalign*}}
\newcommand{\ees}{\end{equation*}}

\newcommand{\eit}{\end{itemize}}

\newcommand{\eo}{\mathbf{e_8}}

\newcommand{\eon}{\mathbf{e_8^{(n)}}}

\newcommand{\esei}{\mathbf{e_6}}
\newcommand{\esn}{\mathbf{e_6^{(n)}}}

\newcommand{\est}{\mathbf{e_7}}
\newcommand{\estn}{\mathbf{e_7^{(n)}}}

\newcommand{\fff}{\mathfrak F}

\newcommand{\fq}{\mathbf{f_4}}
\newcommand{\fqn}{\mathbf{f_4^{(n)}}}

\newcommand{\gd}{\mathbf{g_2}}

\newcommand{\ggo}{\mathbf{\mathfrak g_0}}

\newcommand{\hu}{\mathbf{H}}

\newcommand{\jotn}{\mathbf{J_3^n}}
\newcommand{\jobtn}{\mathbf{\overline J_3^{\raisebox{-2 pt}{\scriptsize \textbf n}}}}

\newcommand{\joto}{\mathbf{J_3^8}}
\newcommand{\jobto}{\mathbf{\overline J_3^{\raisebox{-2 pt}{\scriptsize \textbf 8}}}}
\newcommand{\jp}{\circ}
\newcommand{\lk}{\mathfrak{L}}
\newcommand{\Ll}{\mathbb L}

\newcommand{\lms}{{\bf{\mathcal L_{MS}}}}

\newcommand{\lra}{\leftrightarrow}

\newcommand{\nbf}{\mathbf{n}}

\newcommand{\omv}{\omega^v}
\newcommand{\oms}{\omega^s}
\newcommand{\oo}{\textbf{\large $\mathfrak C$}}

\newcommand{\ql}{\textquotedblleft}
\newcommand{\qr}{\textquotedblright}
\newcommand{\qq}{\mathbf{Q}}

\newcommand{\rr}{\mathbf{R}}

\newcommand{\str}{\text{str}}

\newcommand{\tfo}{T_O}
\newcommand{\tfop}{T_O^\prime}
\newcommand{\tfs}{T_S}
\newcommand{\tfsm}{T_S^-}
\newcommand{\tfsp}{T_S^+}
\newcommand{\tfspm}{T_S^\pm}

\newcommand{\um}{{\scriptstyle \frac12}}

\newcommand{\xd}{x_{P2}}
\newcommand{\xdb}{\bar x_{P2}}
\newcommand{\xpi}{x_{Pi}}

\newcommand{\xt}{x_{P3}}
\newcommand{\xtb}{\bar x_{P3}}
\newcommand{\xu}{x_{P1}}
\newcommand{\xub}{\bar x_{P1}}

\newcommand{\zz}{\mathbb Z}

\numberwithin{equation}{section}
\begin{document}

\begin{titlepage}
\begin{center}

\vskip 3.0cm

{\bf \huge Magic Star and Exceptional Periodicity: \\\vskip 5pt an approach to Quantum Gravity}

\vskip 3.0cm

{\bf \large Piero Truini${}^{1}$, Alessio Marrani${}^{2}$, and Michael Rios${}^{3}$ }

\vskip 40pt

{\it ${}^1$Dipartimento di Fisica and INFN, Universit\`{a} di Genova,\\
Via Dodecaneso 33, I-16146 Genova, Italy}\\ \vskip 5pt

\vskip 40pt

 {\it ${}^2$Museo Storico della Fisica e Centro Studi e Ricerche ``Enrico Fermi'',\\
Via Panisperna 89A, I-00184, Roma, Italy}\\\vskip 5pt

\vskip 40pt

{\it ${}^3$Dyonica ICMQG,\\5151 State University Drive, Los Angeles, CA 90032, USA}\\

\vskip 30pt

\texttt{truini@ge.infn.it},
\texttt{jazzphyzz@gmail.com},
\texttt{mrios@dyonicatech.com}

\end{center}

\vskip 95pt

\begin{center} {\bf ABSTRACT}\\[3ex]\end{center}

We present a periodic infinite chain
of finite generalisations of the exceptional structures,
including the exceptional Lie algebra $\eo$,
the exceptional Jordan algebra (and pair) and the octonions.
We will also argue on the nature of space-time
and indicate how these algebraic structures may
inspire a new way of going beyond
the current knowledge of fundamental physics.

\vskip 45pt

\begin{center}
Presented by P.T. at the \textit{32nd International Colloquium on Group Theoretical Methods in Physics},\\Prague, July 9-13, 2018
\end{center}

\vskip 30pt

\begin{center}
Dedicated to I.E. Segal (1918-1998)  in commemoration of the centenary of his birth
\end{center}




%
\vfill

\end{titlepage}

\newpage \setcounter{page}{1} \numberwithin{equation}{section}

\section{Introduction}
The exceptional Lie algebra $\eo$ is presented as an interplay between orthogonal sub-algebras with their spinors and Jordan pair valued representations of $\adue$ - the complex form of $\mathbf{su(3)}$.
It is {\it exceptionally interesting} for Physics that the exceptional algebras are built out of orthogonal algebras and spinors since the spinors are not only viewed as representations of the orthogonal sub-algebra but they have a non trivial adjoint action among themselves, like fermions, and with the generators of the orthogonal sub-algebra, that can be associated to bosons.
The Jordan structure is clearly exhibited by the {\it magic star}, arising from the projection of the roots of $\eo$ on the plane of an $\adue$.
It is remarkable that the same structure of sub-algebras, spinors and Jordan algebras can be extended to an infinite chain, called {\it Exceptional Periodicity}, of finite dimensional algebras $\eon$ of rank $N=4(n+1)$, $n=1,2,...$. For $n=1$ we have $\eo$, whereas for $n>1$ $\eon$ is not a Lie algebra, but maintains many properties of it except for the Jacobi identity, which fails in the spinorial part. The Jordan structure generalises to a Hermitian matrix algebra structure, dubbed T-algebra, which also maintains many properties of the Jordan algebras, including the content of three scalars, one vector and two Weyl spinors of an orthogonal sub-algebra of $\eon$.
The lack of the Jacobi identity for the algebras in the Exceptional Periodicity implies that these algebras cannot be exponentiated to a group. A discussion on the nature of space-time will indicate how this is not an issue as these algebras assume the role of discrete vertex operator algebras for a theory in which a discrete space-time is created by the interactions in an intuitive way.

\section{From $\eo$ to Exceptional Periodicity}

\begin{figure}[tbp]
\centering
\includegraphics[scale=0.5]{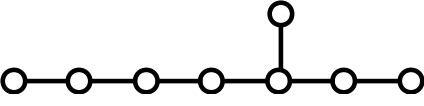}
\caption{The Dynkin diagram of $\eo$}
\label{f:Dynkin}
\end{figure}
\begin{figure}[tbp]
\centering
\includegraphics[scale=0.5]{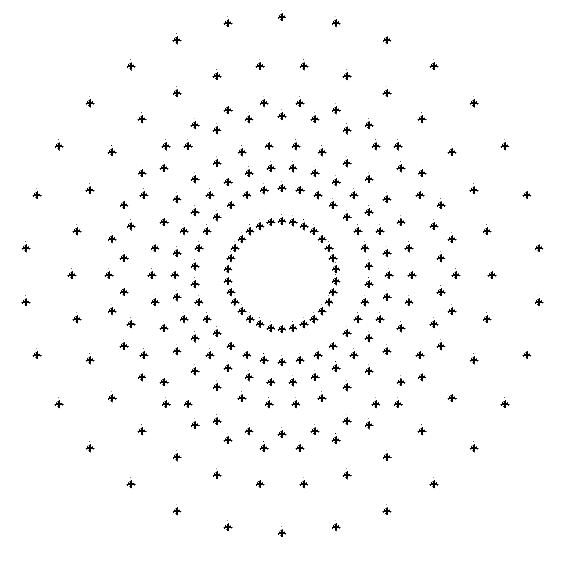}
\caption{The roots of $\eo$ projected on the Coxeter plane}
\label{f:rosone}
\end{figure}

The way $\eo$ is usually presented is through its Dynkin diagram, see Fig. \ref{f:Dynkin},
or through its beautiful projection on the Coxeter plane, see Fig. \ref{f:rosone}.

\begin{figure}[tbp]
\centering
\includegraphics[scale=.5]{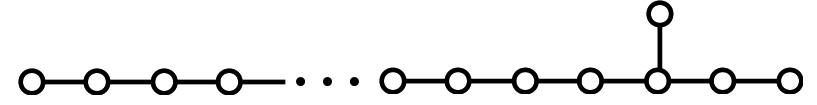}
\caption{The generalised simple roots of $\eon$}
\label{f:Dynkineon}
\end{figure}

The Dynkin diagram of $\eo$ can be obtained from that of $\dset$ by adding a $\dotto$ Weyl spinor:
\bes\label{sroots}
\begin{array}{l}
\phantom{\Delta = \{ k_1-k_2> k_2-k_3>} {\mathbf {d_7}}\\
\Delta = \{\overbrace{\underbrace{k_1-k_2> k_2-k_3> ... > k_6 - k_7} > k_6 + k_7} > -\um(k_1+k_2+...+k_8)\}\\
\phantom{\Delta = \{ k_1-k_2> k_{22}-}{\mathbf {a_6}}
\end{array}
\ees
One gets from these simple roots all the roots of $\dotto$ plus all its 128 spinors to form the root system of $\eo$ (240 roots):
\bea{lll}\label{e8roots}
\pm k_i\pm k_j & 1\le i<j\le 8 & 112\ roots\\
\frac12 (\pm k_1 \pm k_2 \pm k_3 \pm k_4 \pm k_5 \pm k_6 \pm k_7 \pm k_8) &\text{even \# of +} & 128\ roots
\eea
In the same way we obtain the {\it generalised} simple roots of the Exceptional Periodicity of dimension $N=4(n+1)$, $n=1,2,...$.\\
The $\ddn$ spinor that we add to the roots of $\mathbf {d_{N-1}}$ is
$-\um(k_1+k_2+...+k_N)$
that has the right scalar products with the other simple roots but norm (square length) = $n+1$, see Fig. \ref{f:Dynkineon}. We get the set of generalised roots
\bea{llcl}
\pm k_i\pm k_j & 1\le i<j\le N & 2 N(N-1) &\text{roots}\\
\frac12 (\pm k_1 \pm k_2 \pm ... \pm k_N) &\text{even \# of +} & 2^{N-1} &\text{roots}\label{reon}
\eea
They do not form a root system, since they are not invariant under Weyl reflections nor $2 \dfrac{(\alpha ,\beta)}{(\alpha , \alpha)}$ is an integer, for all roots $\alpha, \beta$, due to the norm of the spinors. Note that this extension of the Dynkin diagram of $\eo$ is completely different from the extension giving rise to the Kac-Moody algebras $\mathbf{e_9}$, $\mathbf{e_{10}}$ and $\mathbf{e_{11}}$, \cite{kac}, which are all infinite-dimensional.

\section{The Magic Star for $\eo$}

The beautiful projection on the Coxeter plane does not show the substructures of $\eo$. We now focus on a particular projection of the roots of $\eo$ on a plane, {\it the magic star}, which corresponds to astonishing algebraic properties, introduced in \cite{mukai} and later in \cite{pt1}, with a different perspective involving Jordan Pairs \cite{loos}.

\begin{figure}[tbp]
\centering
\includegraphics{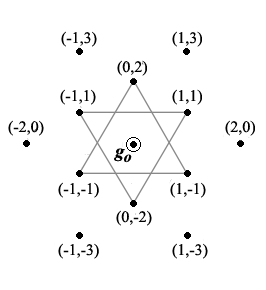}
\caption{Magic star of $\eo$ on the plane of $\adc$}\label{ms0}
\end{figure}

There are four orthogonal sets of roots in $\Phi_8$ corresponding to four $\adue$ root systems. The projection of the $\eo$ roots on any of these 4 planes is the magic star like the one in Fig. \ref{ms0}, which is obtained, for instance, by projecting on the plane of the $\adue$ with roots $\pm (k_i - k_j)$, $i<j = 1,2,3$, dubbed $\adc$, and by arranging the $\eo$ roots according to the scalar products $(r,s)$, where, for each root $\as$, $r:=(\as,k_1-k_2)$ and $s:=(\as,k_1+k_2-2k_3)$.\\
The roots that are projected on the origin $(0,0)$, labeled by $\ggo$, form the root system of $\esei$ which in turn, as we shall show in the next section, can be projected onto a magic star.\\
The real \ql magic\qr of the magic star is in its algebra content, so clearly represented in its picture.

\subsection{Jordan algebras and Jordan Pairs}

\begin{figure}[tbp]
\begin{center}
\includegraphics{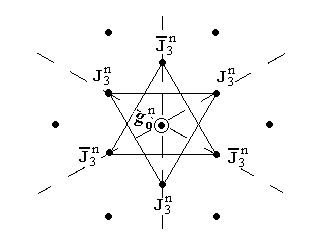}
\end{center}
\caption{A unifying view of the roots of exceptional Lie algebras}
\label{jms}
\end{figure}

A Jordan Pair, \cite{loos}, is just a pair of modules $(J^+, J^-)$ acting on each other, but not on themselves, through a quadratic map $U_{x^\sigma}: J^{-\sigma} \to J^{\sigma}$ and its linearisation $V_{x^\sigma,y^{-\sigma}}: J^{\sigma} \to J^{\sigma}$, where $\sigma = \pm$, $x^\sigma \in J^{\sigma}$, $y^{-\sigma} \in J^{-\sigma}$.\\
It has been proven in \cite{pt1} that three Jordan pairs, pairing Jordan algebras of $3\times 3$ matrices, are at the core of the exceptional Lie algebras as the magic star of Fig. \ref{jms} explicitly shows.

There are three Jordan pairs $(\jotn,\jobtn)$, each lying on an axis, symmetrically with respect to the center of the diagram. Each pair doubles a simple Jordan algebra of rank $3$, $\jotn$, the algebra of $3\times 3$ Hermitian matrices over $\hu$, where $\hu=\rr,\,\cc  ,\,\qq,\,\oo$ for $\nbf=1,2,4,8$ respectively, stands for real, complex, quaternion, octonion algebras.
Exceptional Lie algebras $\fq$, $\esei$, $\est$, $\eo$ are obtained for $\nbf%
=1,2,4,8$, respectively. $\gd$ can be also represented in the same way, with the Jordan algebra reduced to a single element. The Jordan algebras $\jotn$ (and their conjugate $\jobtn$) globally
behave like a $\mathbf{3}$ (and a $\mathbf{\overline{3}}$) dimensional
representation of the outer $\adue$. The algebra denoted by $\ggo$ in the center (plus the Cartan generator associated with the
axis along which the pair lies) is the algebra of the
automorphism group of the Jordan Pair  $(\jotn,\jobtn)$,
the structure group of the corresponding Jordan algebra $\jotn$.\\
\begin{remark}
The quadratic formulation of Jordan algebras and Jordan pairs particularly suits our purposes of finding Jordan algebras, based on a symmetric product, having only commutators in our hands. If one looks at Fig. \ref{jms} and focuses on just one dot where a Jordan algebra sits then the commutator of two elements there is obviously $0$, since the sum of two roots projected on that dot is certainly not a root.
If one considers however a pair $(x^+,x^-)$ in $(\jotn,\jobtn)$ then $[x^+,x^-]$ is either zero or is in $\ggo\oplus\cc$, therefore $[[x^+,x^-],y^+]$ is in $\jotn$. Moreover the Jacobi identity applied to these elements reads:
$$[[x^+,x^-],y^+] = [[y^+,x^-],x^+] \qquad symmetric\ in \ x^+,y^+$$
being $[x^+,y^+]=0$. We get a trilinear product like $V_{x^+,x^-}z^+$, which is symmetric by definition in $x^+,y^+$. The Jordan Pair axioms follow directly from the Jacobi identity, \cite{mc}.
\end{remark}

In fact Jordan pairs are strongly related to the Tits-Kantor-Koecher construction of Lie Algebras \cite{tits1}-\nocite{kantor1}\cite{koecher1}:
\begin{equation}
\lk = J \oplus \str(J) \oplus \bar{J} \label{tkk}
\end{equation}
where $(J,\bar J)$ is a Jordan pair, with trilinear product
$V_{x^\sigma , y^{-\sigma}} z^\sigma = [[x^\sigma , y^{-\sigma}], z^\sigma]$, $str(J)$ is the structure algebra of $J$ which is also the algebra of derivations of the Jordan pair $(J, \bar J)$.

 In the case of (complex) exceptional Lie algebras this construction applies to $\est$, with $J = \joto$, the 27-dimensional exceptional Jordan algebra of $3 \times 3$ Hermitian matrices over the octonions, and $\str(J) = \esei \oplus \cc$ ($\mathbf{C}$, the complex field). An $\est$ sub-algebra is clearly exhibited by the magic star by taking the central $\esei$ and a $(\joto, \jobto)$ pair on opposite side with respect to the center. The fundamental {\bf 56} dimensional representation of this $\est$ is also easily shown: it is a Freudenthal triple system, \cite{freu} \cite{hel}, containing a $\joto$, a $\jobto$ and two dots from $\adue$, lying in the same direction of $\est$, parallel to it.\\

Out of this construction Kantor drew the conclusion that there are no Jordan Algebras: there are only Lie algebras. On the other hand McCrimmon said: {\it if you open up a Lie algebra and look inside, 9 times out of 10 there is a Jordan algebra (or pair) which makes it tick.}
\section{Exceptional periodicity}
We now introduce the concept of Exceptional Periodicity, a chain of finite structures of dimensions $4(n+1)$, with $n=1,2,...$ that extend $\eo$, the case $n=1$, both as a set of roots and as an algebra.
\subsection{Exceptional periodicity generalised roots}
We force the definition of root system to include what we call {\it generalised} roots not obeying the symmetry by Weyl reflection, nor the fact that $2\dfrac{(\alpha,\beta)}{(\alpha,\alpha)}$ be integer for all roots $\alpha$, $\beta$.\\
For any $n=1,2,...$ we introduce $N=4(n+1)$ and define the generalised roots of $\eon$ as in \eqref{reon}.
This is a root system only in the case $n=1$, being $\eo^{(1)}=\eo$.
The {\it generalised} roots of $\fqn$, $\esn$, $\estn$ can also be  obtained in a similar fashion.\\
For any $N$ the sets of {\it generalised} roots form a magic star as in Fig. \ref{ms0}, once projected on the plane spanned by $k_1-k_2$ and $k_1+k_2-2k_3$.

Similarly to the case of $\eo$, the roots \footnote{By abuse of definition we shall often say {\it root}, for short, instead of
{\it generalised root}.} of $\esn$ lie in the center of the magic star of $\eon$ and those of $\estn$ can be writen as $\estn=\esn\oplus T_{(r,s)}\oplus T_{(-r,-s)}$, for a fixed pair $(r,s)\in \{(1,1),(-1,1),(0,-2)\}$, where $T_{(r,s)}$ is the $(r,s)$ set of roots $\as$ such that $(\as,k_1-k_2)=r$ and $(\as,k_1+k_2-2k_3)=s$.\\

We denote by $\Phi$ the set of roots of $\eon$ in the Euclidean vector space of dimension $N=4(n+1)$, $n=1,2,...$ and by $\Phi_O$ and $\Phi_S$ the following subsets of $\Phi$:
\be
\Phi_O = \{ (\pm k_i \pm  k_j) \in \Phi \}
\qquad
\Phi_S =  \{ \frac{1}{2} (\pm k_1 \pm k_2 \pm ... \pm k_N)  \in \Phi \}
\ee

\begin{remark}\label{r:dd} Notice that $\Phi_O$ is the root system of  $\ddn$. The set of generalised roots is closed under the Weyl reflections by all roots if and only if $n=1$. Nonetheless the set $\Delta$
\be
\Delta = \{ k_1-k_2> k_2-k_3> ... > k_{N-2} - k_{N-1} > k_{N-2} + k_{N-1}  > -\um(k_1+k_2+...+k_N)\}
\ee
is a set of {\bf generalised  simple roots}, by which we mean:
\bit
\item[i)] $\Delta$ is a basis of the Euclidean space $V$ of finite dimension $N$;
\item[ii)] every root $\beta$ can be written as a linear combination of roots of $\Delta$ with all positive or all negative integer coefficients: $\beta = \sum \ell_i \alpha_i$ with $\ell_i \ge 0$ or $\ell_i\le 0$ for all $i$.
\eit
\end{remark}

\subsection{The $\lms$ algebra for EP}\label{s:lmsEP}

We define the $\lms$ algebra ($\lms = \eon$ in this paper) by extending the construction used for Lie algebras, \cite{carter}-\nocite{hum}\cite{graaf}.\\
We give $\lms$ an algebra structure of rank $N$ over a field extension $\fff$ of the rational integers $\zz$ in the following way
\footnote{Specifically, we will take $\fff$ to be the complex field $\cc$.}
:
\bit
\item[a)]  we select the set of simple generalised ordered roots $\Delta = \{\alpha_1 , ... ,\alpha_N\}$ of $\Phi$
\item[b)] we select a basis $\{ h_1 ,...,h_N\}$ of the $N$-dimensional vector space $H$ over $\fff$ and set $h_\alpha = \sum_{i=1}^N c_i h_i$ for each $\alpha  \in \Phi$ such that $\alpha = \sum_{i=1}^N c_i \alpha_i$
\item[c)] we associate to each $\alpha  \in \Phi$ a one-dimensional vector space $L_\alpha$ over $\fff$ spanned by $x_\alpha$
\item[d)] we define $\lms = H \bigoplus_{\alpha \in \Phi} {L_\alpha}$ as a vector space over $\fff$
\item[e)] we give $\lms$ an algebraic structure by defining the following multiplication on the basis
$\blms = \{ h_1 ,...,h_N\} \cup \{x_\alpha \ | \ \alpha \in \Phi\}$, extended by linearity to a bilinear multiplication $\lms\times \lms\to \lms$:
	\be\begin{array}{ll}
	&[h_i,h_j] = 0 \ , \ 1\le i, j  \le N \\
	&[h_i , x_\alpha] = - [x_\alpha , h_i] = (\alpha, \alpha_i )\, x_\alpha \ , \ 1\le i \le N \ , \ \alpha \in \Phi \\
	&[x_\alpha, x_{-\alpha} ] = - h_\alpha\\
	&[x_\alpha,x_\beta] = 0 \ \text{for } \alpha, \beta \in \Phi \ \text{such that } \alpha + \beta \notin			 	\Phi \ \text{and } \alpha \ne - \beta\\
	&[x_\alpha,x_\beta] = \varepsilon (\alpha , \beta)\,  x_{\alpha+\beta}\ \text{for } \alpha , \beta \in \Phi \ \text{such that }  \alpha+ \beta \in \Phi\\
	\end{array} \label{comrel}\ee
\eit
where $\varepsilon (\alpha , \beta)$ is the {\it asymmetry function}, \cite{kac}, extended to the roots of $\eon$ as follows:\\
\begin{definition} Let $\Ll$ denote the lattice of all linear combinations of the simple generalised roots with integer coefficients. The asymmetry function $\varepsilon (\alpha , \beta) : \ \Ll \times \Ll \to \{-1,1\}$ is defined by:
\be\label{epsdef}
\varepsilon (\alpha , \beta) = \prod_{i,j=1}^N \varepsilon (\alpha_i , \alpha_j)^{\ell_i m_j} \quad \text{for } \alpha = \sum_{i=1}^N \ell_i\alpha_i \ ,\ \beta = \sum_{j=1}^N m_j \alpha_j
\ee
where $\alpha_i , \alpha_j \in \Delta$ and
\be
\varepsilon (\alpha_i , \alpha_j) = \left\{
\begin{array}{ll}
-1 & \text{if } i=j\\ \\
-1 & \text{if } \alpha_i + \alpha_j  \text{ is a root and } \alpha_i < \alpha_j\\ \\
+ 1 & \text{otherwise}
\end{array}
\right.
\ee
\end{definition}

\begin{remark}
For $n>1$ the adjoint action $\adj_x:y \to [x,y]$ is a derivation of the algebra $\lms$ (and hence $\exp(\zeta \adj_x)$ is an automorphism of $\lms$) if and only if $x\in \ddn$.\\
This means in particular that the Jacobi identity does not hold for $\eon$, except for $n=1$. We know that the exceptional chain of Lie algebras stops at $\eo$, unless we jump to infinite-dimensions. The algebras $\eon$ are an infinite chain of finite dimensional extensions of $\eo$. The price to be paid is to give up Jacobi. This means that the generators of $\eon$, for $n>1$, cannot be related to the infinitesimal action of a group. We can disregard this problem, if we accept that the physical transformations are discrete repeated actions of the generators and that space-time emerges from the interactions, thus being discrete as well. We will come back to this point in section 5.
\end{remark}

\subsection{Exceptional T-algebras}

Let us concentrate on the set of roots $T_{(r,s)}$ on a particular tip $(r,s)$ of the magic star in Fig. \ref{ms0}, for instance on $T_{(1,1)}$ that we simply denote by $T$.\\
We will use $T$ to denote both the set of roots and the set of elements in $\lms$ associated to those roots. An element of $T$ is an $\fff$-linear combination of $x_\as, x_\beta, ...$ for $\as,\bb,...$ in $T_{(1,1)}$. The roots of $T_{(1,1)}$ are:
\bea{llll}
- k_2 -  k_3 \ , \  k_1 \pm k_i & i=4,... , N &  & 2N-5\\
\frac{1}{2} (k_1 - k_2 - k_3 \pm k_4 \pm ... \pm k_N) & \text{even \# of +} &  & 2^{N-4}
\eea

In \cite{trm1} we gave $T$ an algebraic structure with a symmetric product, thus mimicking the case $n=1$ when $T$ is a Jordan algebra.\\
We define $\xu$, $\xd$ and $\xt$ as the elements of $\lms$ in $T$ associated to the roots $\rho_1:=k_1+ k_N$, $\rho_2:=k_1- k_N$ and $\rho_3:=-k_2-k_3$:
\be\label{notp}
\xu \lra \rho_1:=k_1+ k_N \ ; \ \xd \lra \rho_2:=k_1- k_N \ ; \ \xt \lra \rho_3:=-k_2 - k_3
\ee
They are left invariant by the Lie sub-algebra $\mathbf{d_{N-4}}=\dqn$, whose roots are $\pm k_i \pm k_j \ , \ 4\le i<j \le N-1$.\\
We denote by $\tfo$ the set of roots in $T\cap \Phi_O$,  by $\tfop$ the set of roots in $\tfo$ that are not $\rho_1,\rho_2,\rho_3$ and by $\tfs$ the set of roots in $T\cap \Phi_S$. In the case we are considering, where $T=T_{(1,1)}$ we have $\tfop = \{ k_1\pm k_j\, ,\ j=4,...,N-1\}$ and $\tfs= \{ \frac{1}{2} (k_1 - k_2 - k_3 \pm k_4 \pm ... \pm k_N)\}$, even $\#$ of $+$. We further split $\tfs$ into $\tfsp = \{ \frac{1}{2} (k_1 - k_2 - k_3 \pm k_4 \pm ... + k_N)\}$ and $\tfsm= \{ \frac{1}{2} (k_1 - k_2 - k_3 \pm k_4 \pm ... - k_N)\}$. Then $v\in \tfop$ is an $8n$-dimensional vector and $s^\pm$ are $2^{4n-1}$-dimensional chiral spinors of $\dqn$.

We write a generic element $x$ of $T$ as  $x= \sum_{i=1}^3 \lambda_i \xpi + \, x_v\, +\, x_{s^+}+x_{s^-}$ where
\be
x_v = \sum_{\as \in \tfop} \lambda_\as^v x_\as
\qquad
x_{s^\pm} = \sum_{\as \in \tfspm} \lambda_\as^{s^\pm} x_\as
\ee

We view $\lambda_\as^v$ as a coordinate of the vector $\lambda^v$ and $\lambda_\as^{s^\pm}$ as a coordinate of the spinor $\lambda^{s^\pm}$; we denote by  $\bar\lambda^v$ ($\bar\lambda^{s^\pm}$) the vector ( spinor) in the dual space with respect to the bilinear form $\omv$ ($\oms$), that we are not going to specify here, and view $x$ as a $3\times 3$ Hermitian matrix:

\be\label{matrix}
\left(
\begin{array}{lll}
\lambda_1 &\lambda_v &\bar  \lambda_{s^+}\\
\bar \lambda_v &\lambda_2 &\lambda_{s^-}\\
\lambda_{s^+} &\bar  \lambda_{s^-} &\lambda_3
\end{array}
\right)
\ee
 whose entries have the following $\fff$-dimensions:
\bit
\item 1 for the {\it scalar} diagonal elements $\lambda_1,\lambda_2,\lambda_3$;
\item $8n$ for the {\it vector} $\lambda_v$;
\item $2^{4n-1}$ for the {\it chiral spinors} $\lambda_{s^\pm}$.
\eit

We see that only for $n=1$ the dimension of the vector and the spinors is the same, whereas for $n>1$ the entries in the $(12), (21)$ position have different dimension than those in the $(31),(13),(23),(32)$ position. Nevertheless we now define a symmetric product of the elements in $T$, which then becomes a generalization of the Jordan algebra $\joto$ in a very precise sense. This type of generalization of the Jordan Algebra is known in the literature, \cite{vin}, where it is called $T-algebra$. In our case, since we are generalizing the Exceptional Jordan algebra we call it
{\it Exceptional T-algebra}.

We denote by $I$ the element $I:=\xu+\xd+\xt$ and by $I^-$ the element $I^-:=-\xub-\xdb-\xtb$ of $\bar T:= T_{(-1,-1)}$, where $\xub$, $\xdb$ and $\xtb$ are associated to the roots $-k_1- k_N$, $-k_1+ k_N$ and $k_2+k_3$.\\
We give $T$ an algebraic structure by introducing the symmetric product, \cite{trm1}:
\be\label{jprod}
x\jp y := \frac12 [[x,I^-],y] \quad ,\quad x,y\in T
\ee

We introduced in \cite{trm1} an (associative)  trace,  a generalized norm, rank 1, rank 2, rank 3 elements just like in  Jordan algebra, thus completing the generalization of the magic star algebraic content of $\eo$ to the whole chain of the exceptional periodicity.

\section{A model for an expanding Universe}\label{s:eu}

We now discuss a possible application of $\eon$ to Quantum Gravity. In particular we view $\eon$ as a code - a set of rules - for a Universe in which space-time is created and expands.
\subsection{$\eon$ in a physical perspective}

Exceptional groups and algebras and the underlying non-associative algebra of the octonions have attracted the attention of many theoretical physicists since the pioneering work of F. G$\ddot{\text{u}}$rsey, \cite{unif}, see for instance \cite{MESGT}$\div$\nocite{Het}\nocite{Sati-1}\nocite{12}\nocite{13}\cite{14}.

\begin{figure}[htbp]
\begin{center}
\includegraphics{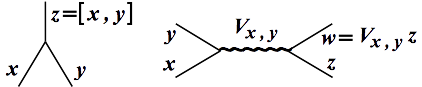}
\end{center}
\caption{Building blocks of the interactions and an elementary scattering process}
\label{int1}
\end{figure}

If we look inside $\eo$ we see four orthogonal $\mathbf {su(3)}$'s - $\adue$'s in their complex form. Beside $\adc$, for color we see
$\adf$ that we associate to flavor degrees of freedom and two more, in the center of the {\it flavor} magic star that lies in the center of the {\it colored} magic star: in this sense {\it at the core of the core of the magic star}. It is natural to associate these two $\adue$'s to gravity, but in the context of what theory?\\
It seems natural to think of $\eo$, and in general $\eon$, as a model for a theory of everything, including space-time, whose classical concept needs to be reviewed in the light of  the incompatibility of quantum mechanics and general relativity at the Planck scale, which makes it impossible to test, in particular, if space-time is a continuum.\\
An important feature of space-time is that it is {\em dynamical} and related to matter, as Einstein taught us in his theory of general relativity.
The Big Bang, for instance, is not a blast in empty space. Physicists do not think there was a space and the Big Bang happened in it: Big Bang was a blast of space-time itself, as well as matter and energy. If two galaxies withdraw from each other it simply means that space is being created between them.\\
There is only one way physicists know to reproduce the process of creation: make particles interact. We state that:
\begin{center}{\it there is no way of defining space-time without a preliminary concept of interaction}.
\end{center}
Stated differently, a universe of non-interacting particles has no space-time: there is no physical quantity that can relate one particle to another. Our basic principle implies that we have to start with a model of interactions, consistent with the present observations, and deduce from  it what space-time is. This is the way we look at $\eo$ and its extension to exceptional periodicity.\\
We are used to start from space-time because our point of view is that of an observer (who measures things in space-time). However, if we want to describe extreme situations like at the Planck scale or right after the Big Bang, we can no longer give any meaning to the concept of observer.\\
We notice that all fundamental interactions look similar at short distances. Their basic structure is very simple: it involves only three entities, like the product in an algebra. The first step in our approach is to define objects and {\it elementary interactions}, with the hypothesis in mind, similar to the Bethe \textit{Ansatz} in integrable models, that every interaction is made of elementary interactions. This hypothesis gives the interactions a tree structure, thus opening the way for a description of scattering amplitudes in terms of associahedra or permutahedra or a generalisation thereof, to account for the expansion of space-time.\\
An elementary - or fundamental - interaction may be defined as the interaction between $x$ and $y$ in $\lms$ to produce the outcome $z$ in $\lms$. It is represented by $(x,y \rightarrow z) \leftrightarrow [x,y] = z$, see Fig. \ref{int1}.\\

\subsection{Emergent Space-time: Outline}

In order to build a model where space-time emerges from the interactions we must start from scratch, namely from the initial conditions of a Big Bang. The minimal initial conditions are those in which the least number of generators is taken, namely the generators corresponding to a minimal set of roots whose linear combination generates all the roots. Since the universe is bounded we assign a superposition of opposite 3-momenta $\vec p$ and $-\vec p$, as for a quantum particle in a box, to all initial generators which are then allowed to interact among themselves. This is interpreted, having in mind locality, by the fact that the initial particles are all at the same point, even though there is no geometry, no singularity and actually no point of an \textit{a priori} space. The initial momenta are assigned according to a specific projection on 3D of the root space. The initial particles are assumed to be on-shell massless. Energy-momentum conservation is assumed from there on.\\
Each interaction has two effects: produce a new particle, according to the commutation rules, and create two new points, $\pm {\vec p}/E$ apart from the point of the interaction, where $(E,{\vec p})$ is the 4-momentum of the produced particle.\\
The quantum nature of the model is imposed by saying that each particle has a chance to interact, thus producing new particles, but also a chance not to interact. In this latter case the particle only creates two new points, shifted with respect to the previous position by $\pm{\vec p}/E$, $(E,\vec p)$ being its 4-momentum. The interaction probability amplitudes are given by the structure constants of $\eon$.\\
This is the first stage of interactions. We see the outcome and pass to a second stage and so on. We can intuitively associate a {\it cosmological discrete (quantum of) time} with each stage of interactions.\\
What emerges is a {\it quantum field}:  we identify the field itself with a generator of $\eon$, that spreads as space-time is created as a quantum wave.\\
It is obvious that the space-time emerging in the approach outlined here is dynamical, finite and discrete, being the outcome of a countable number of interactions among a finite number of objects. This is in agreement with the two cutoffs coming from our current knowledge of Physics: the background radiation temperature (finiteness) and Planck length (discreteness). The granularity of space-time implies that the velocity of propagation of the interaction is also discrete and finite. If the distance traveled from one level of interactions and the next one is 1 Planck length and the time interval is 1 Planck time then the maximum speed of propagation is the speed of light. This quantum model is intrinsically relativistic. Our approach leads to a finite model by construction, with the continuum limit as a macroscopic approximation.\\
We also have that all the infinities or continuities of the standard theories are not present: we have no symmetry Lie groups, just algebras. This is the reason why we look for extensions of $\eo$, like Exceptional Periodicity, which do not extend the Lie group $\mathbf{E_8}$ because of lack of the Jacobi identity. A good reason for extending $\eo$ is to have a larger amount of dark matter degrees of freedom, as requested in modern cosmology, \cite{dm1} \cite{dm2}.\\
A sentence by Segal gives us the appropriate conclusion to this argumentation:

\begin{center}
{\it
To deal with all these issues is a tall order,\\
necessarily at best a matter of successive approximations;\\
but it is useful, and is quite possibly essential,
for us from time to time\\ to view fundamental physics
with maximal perspective.}  I.E.Segal, \cite{segal}
\end{center}

\section*{References}

\end{document}